\documentclass[aps,11pt,prd,longbibliography,notitlepage,nofootinbib]{revtex4-1}
\usepackage{tikz}
\usepackage{feynmp-auto}
\usepackage{amsmath}
\usepackage{todonotes}
\usepackage[utf8]{inputenc}
\usepackage[T1]{fontenc} 
\usepackage{array}
\usepackage{bbm}
\usepackage{soul}	
\usepackage{xcolor}

\usepackage{epsfig}
\usepackage{color}
\usepackage{slashed}
\usepackage{comment}
\usepackage{epstopdf}
\usepackage{xspace}
\usepackage{mathrsfs}
\usepackage[euler]{textgreek}
\usepackage{amsmath}
\usepackage{amsthm}
\usepackage{amsfonts}
\usepackage{amssymb}
\usepackage{graphicx}
\usepackage[font={small}]{caption}
\usepackage{subcaption}
\usepackage{float}
\usepackage{multirow}
\usepackage[hang, flushmargin,bottom]{footmisc} 
\usepackage{placeins}
\usepackage{enumitem}
\usepackage{slashed}
\usepackage{bbm}
\usepackage{appendix}
\usepackage{tabularx}
\usepackage{siunitx}

\usepackage{titlesec}
\titleformat{\chapter}[display]
  {\normalfont\LARGE\bfseries}
  {\chaptertitlename\ \thechapter}{5pt}{\LARGE}
  \titlespacing*{\chapter}{0pt}{-20pt}{35pt}
\usepackage{bigstrut}
\usepackage{pst-node}
\usepackage{pstricks}
\usepackage{physics}
\usepackage{mathtools}
\usepackage{setspace}
\usepackage{fancyhdr}
\usepackage[makeroom]{cancel}
\usepackage{feyn}
\newcommand{\be}{\begin{equation}}
\newcommand{\ee}{\end{equation}}
\newcommand{\bes}{\begin{equation*}}
\newcommand{\ees}{\end{equation*}}

\usepackage{hyperref}
\hypersetup{%
  colorlinks = true,
  linkcolor  = blue,
  citecolor  = blue,
  urlcolor   = blue,
}
\usepackage{soul}
\usepackage{xpatch}
\makeatletter
\xpretocmd{\todo}{\@bsphack}{}{}
\xapptocmd{\todo}{\@esphack}{}{}
\makeatother

\newcommand{\beq}{\begin{equation}}
\newcommand{\eeq}{\end{equation}}

\usepackage{tikz}

\DeclareRobustCommand{\swatch}[1]{\tikz[baseline=-0.6ex]\node[fill=#1,shape=rectangle,draw=black,thick,minimum width=5mm,rounded corners=0.5pt](){};}

\newcommand{\MET}{\ensuremath{p_T^\mathrm{miss}}\xspace}
\newcommand{\gB}{\ensuremath{g_\mathrm{B}}\xspace}
\newcommand{\UB}{\ensuremath{U(1)_\mathrm{B}}\xspace}
\newcommand{\ZB}{\ensuremath{Z_\mathrm{B}}\xspace}
\newcommand{\HB}{\ensuremath{h_\mathrm{B}}\xspace}

\newcommand{\MZB}{\ensuremath{M_{\ZB}}\xspace}

\newcommand{\Mchi}{\ensuremath{M_{\chi}\xspace}}
\newcommand{\Mphi}{\ensuremath{M_{\phi}\xspace}}
\newcommand{\MPsi}{\ensuremath{M_{\Psi}\xspace}}
\newcommand{\tB}{\ensuremath{\theta_\mathrm{B}}\xspace}
\newcommand{\stB}{\ensuremath{\sin\theta_\mathrm{B}}\xspace}
\newcommand{\ctB}{\ensuremath{\cos\theta_\mathrm{B}}\xspace}
\newcommand{\ttbar}{\ensuremath{t\bar{t}}\xspace}
\newcommand{\GeV}{\text{GeV}}
\newcommand{\TeV}{\text{TeV}}

\newcommand{\herwig}{H\protect\scalebox{0.8}{ERWIG}\xspace}
\newcommand{\rivet}{R\protect\scalebox{0.8}{IVET}\xspace}

\newcommand{\contur}{\textsc{Contur}\xspace}
\newcommand{\spey}{\textsc{Spey}\xspace}
\newcommand{\madgraph}{\textsc{Madgraph5}\xspace}

\definecolor{green}{HTML}{008000}
\definecolor{goldenrod}{HTML}{DAA520}
\definecolor{magenta}{HTML}{FF00FF}
\definecolor{silver}{HTML}{C0C0C0}
\definecolor{indigo}{HTML}{4B0082}
\definecolor{skyblue}{HTML}{87CEEB}
\definecolor{darkgoldenrod}{HTML}{B8860B}
\definecolor{orange}{HTML}{FFA500}
\definecolor{yellow}{HTML}{FFFF00}
\definecolor{saddlebrown}{HTML}{8B4513}
\definecolor{blue}{HTML}{0000FF}
\definecolor{turquoise}{HTML}{40E0D0}
\definecolor{yellow}{HTML}{FFFF00}
\definecolor{white}{HTML}{FFFFFF}
\definecolor{whitesmoke}{HTML}{F5F5F5}
\definecolor{hotpink}{HTML}{FF69B4}

\newcommand{\myComment}[1]{}

\newcommand{\Psibar}{\overline{\Psi^-}}
\begin{document}
\title{\LARGE{Local Baryon Number at the LHC}}
\author{Jon Butterworth$^{1}$, Hridoy Debnath$^{2}$, Joseph Egan$^{1}$, Pavel Fileviez P{\'e}rez$^{2}$}
\affiliation{
$^{1}$Department of Physics and Astronomy, University College London, Gower St., London, WC1E 6BT, UK \\
$^{2}$Physics Department and Center for Education and Research in Cosmology and Astrophysics (CERCA), Case Western Reserve University, Cleveland, OH 44106, USA }
\email{j.butterworth@ucl.ac.uk, hxd253@case.edu, joe.egan.23@ucl.ac.uk, pxf112@case.edu}
\date{\today}
\begin{abstract}
  The minimal theory in which baryon number is spontaneously broken at the low scale predicts new fermions, one of which is a
  dark matter candidate, from gauge anomaly cancellation.
  We discuss the production mechanisms and decays of these new fermions, which include channels with multi-leptons,
  and channels with long-lived charged fermions that can give rise to
  exotic signatures with  ``kinked'' tracks at the Large Hadron Collider. 
  We evaluate the contraints on the theory from current LHC searches and measurements, and briefly comment on the excess
  in top pair production at threshold recently reported by CMS. We also
  discuss predictions for the $h \to \gamma \ZB$ decay, where $h$ is the SM-like Higgs and \ZB is
  the new gauge boson associated with baryon number.
\end{abstract}
\maketitle

\section{INTRODUCTION}
After the discovery of the Brout-Englert-Higgs boson at the Large Hadron Collider (LHC)~\cite{ATLAS:2012yve,CMS:2012qbp,CMS:2013btf},
the Standard Model (SM) of Particle Physics stands as one of the most successful ever theories in describing nature.
The SM precisely explains how quarks and leptons interact via the electromagnetic, weak, and strong gauge forces.
In this framework, massive fields acquire mass through the Higgs mechanism, with Higgs boson decays aligning with SM
predictions, consistent with increasingly precise data from the LHC~\cite{CMS:2022dwd,ATLAS:2024fkg}.
As a gauge theory, the SM is built upon the $SU(3)_C\otimes SU(2)_L \otimes U(1)_Y$ gauge group, where lepton and baryon
numbers emerge as accidental global symmetries at the classical level. These global symmetries play a crucial role in neutrino physics and
cosmology.

Extensions to the SM in which the global symmetry associated with baryon number is promoted to a local gauge symmetry
are well-motivated for several
reasons~\cite{FileviezPerez:2010gw,FileviezPerez:2011pt,Duerr:2013dza,FileviezPerez:2014lnj,FileviezPerez:2024fzc},
one of which is that they predict a candidate for dark matter (DM), as studied in detail in Ref.~\cite{Butterworth:2024eyr}.
This DM candidate is one of several new fermions required to cancel the triangle gauge anomalies which would otherwise be
introduced along with the new \UB gauge group.
In Ref.~\cite{Butterworth:2024eyr} some phenomenological aspects of these theories were
studied, under the assumptions that the additional fermions have masses high enough that they play no direct phenomenological
role. All new fermions in this theory acquire mass from the $U(1)_B$ breaking scale and in the majority of cosmologically viable scenarios their masses should be below a few TeV, potentially within reach of colliders.
In this article, we investigate in detail the phenomenological aspects of these fermions needed for anomaly cancellation
in the minimal theory based on local baryon number~\cite{FileviezPerez:2024fzc}.
\begin{figure}[t]
    \centering
    \begin{subfigure}[b]{0.45\textwidth}
        \centering
        \includegraphics[width=\textwidth]{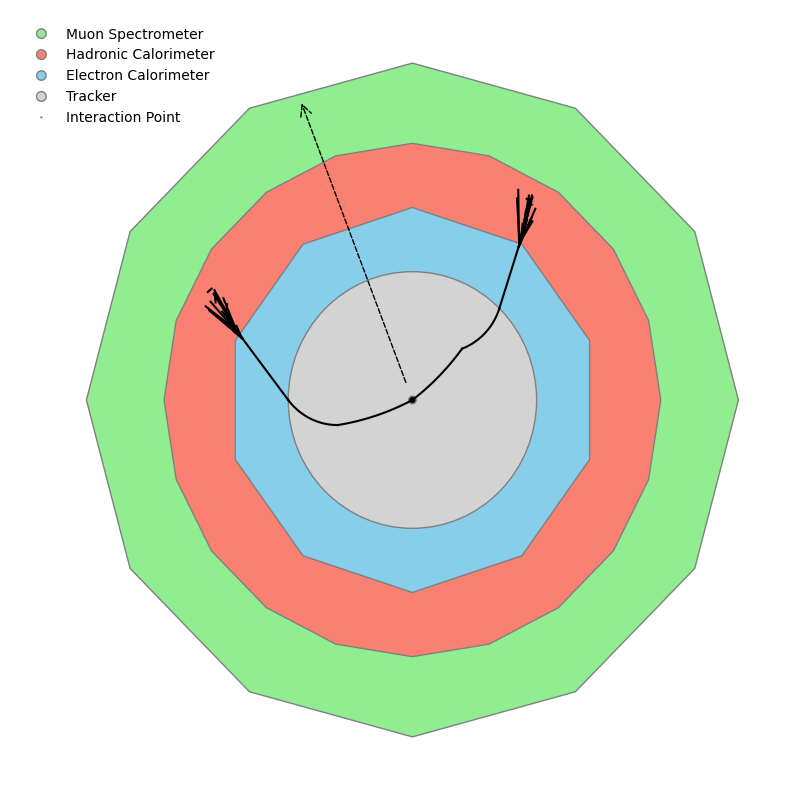}
        \caption{$pp \to \rho^+ \rho^- \rightarrow \rho^0 \rho^0 \pi^+ \pi^-$}
        \label{subfig:rhoMrhoMbar}
    \end{subfigure}  
    \begin{subfigure}[b]{0.45\textwidth}
        \centering
        \includegraphics[width=\textwidth]{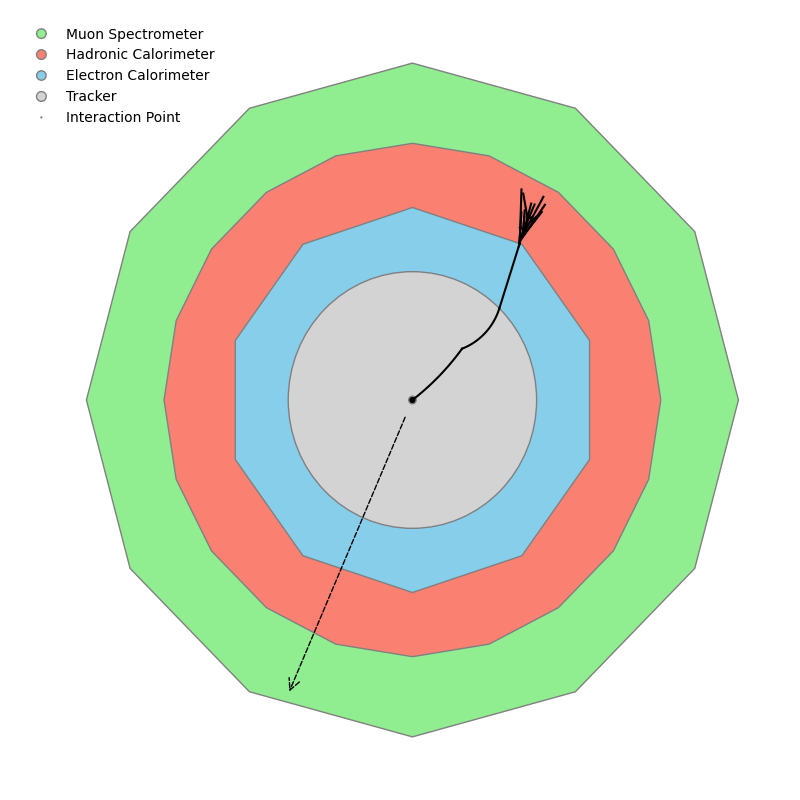}
        \caption{$pp \to \rho^\pm \rho^0 \rightarrow \rho^0 \rho^0 \pi^\pm$}
        \label{subfig:rhoMrhoZ}
    \end{subfigure}          
    \begin{subfigure}[b]{0.45\textwidth}
        \centering
        \includegraphics[width=\textwidth]{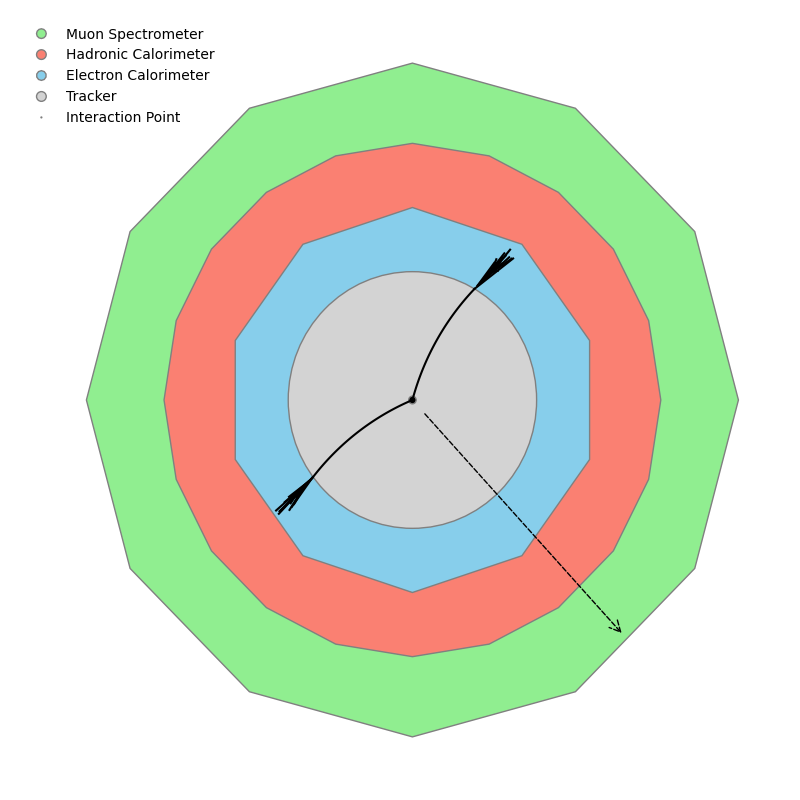}
        \caption{$pp \to \Psi^+ \Psi^- \rightarrow e^+ e^- \phi \phi^*$}
        \label{subfig:psipsi2ee}
    \end{subfigure} 
    \begin{subfigure}[b]{0.45\textwidth}
        \centering
        \includegraphics[width=\textwidth]{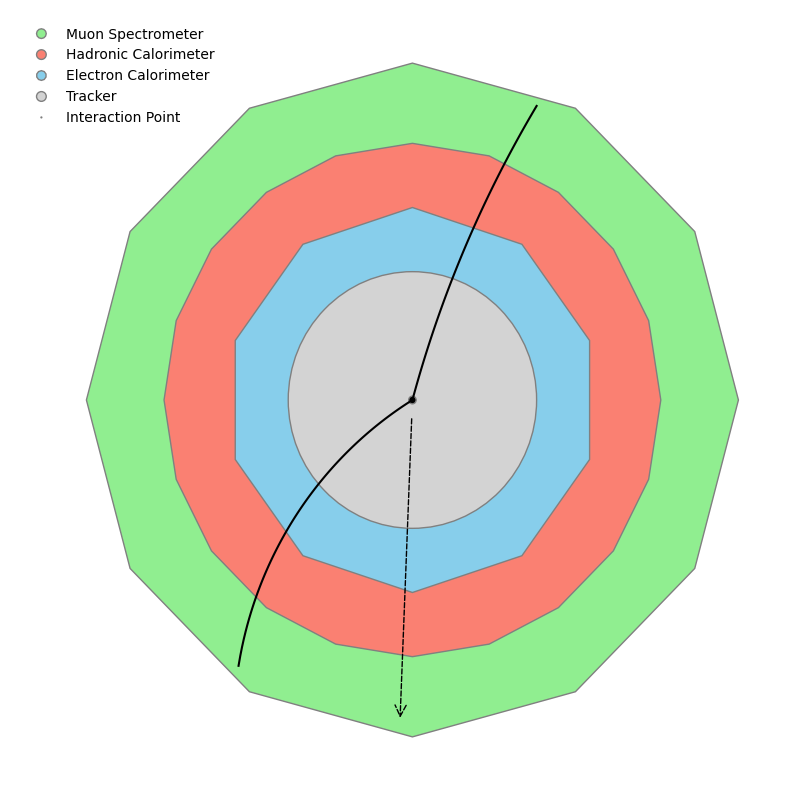}
        \caption{$pp \to \Psi^+ \Psi^- \rightarrow \mu^+ \mu^- \phi \phi^*$}
        \label{subfig:psipsi2mumu}
    \end{subfigure} 
    \caption{Schematics of signatures from the production and decays of (a) $\rho^+ \rho^-$, (b) $\rho^0 \rho^\pm$, (c) $\Psi^+ \Psi^- \to e^+ e^- \phi \phi^*$, and (d) $\Psi^- \Psi^+ \to \mu^+ \mu^- \phi \phi^*$
      at the LHC. $\Psi^- \Psi^+ \to \tau^+ \tau^- \phi \phi^*$ is also possible (not shown).
      The dashed line indicates the summed missing transverse momentum arising from one or more DM
      candidates exiting the detector undetected. The soft charged pion is shown being produced at the $\rho^\pm$ decay point
      and stopping in the hadronic calorimeter, the $e^\pm$ are shown stopping in the electromagnetic
      calorimeter, while the muons are detected in the muon chamber.}
    \label{fig:detector_signatures}
 \end{figure}

In the theory proposed in Ref.~\cite{FileviezPerez:2024fzc} the baryonic anomalies are cancelled with only four new
fermionic representations.
These new fermions acquire mass once the local baryon number is spontaneously broken. This theory predicts two neutral Majorana fermions, $\chi^0$ and $\rho^0$, the lightest of which can be a DM candidate,
and two charged fermions, $\Psi^-$ and $\rho^-$.
One of the charged fermionic fields, $\rho^-$, is long-lived and can give rise to striking signatures at colliders.
The other decays to a SM fermion and missing energy.
In this article, we study in detail all the collider signatures of the theory and point out the regions of parameter space that
are already ruled out by LHC searches and measurements. We find the following distinctive signatures, shown
schematically in Fig.~\ref{fig:detector_signatures}:
\begin{itemize}
\item Two charged tracks, Fig.~\ref{subfig:rhoMrhoMbar}:
  The $pp \to \rho^+ \rho^-\to \rho^0 \rho^0 \pi^+ \pi^-$ channel gives rise
  to two ``kinked'' charged tracks because the $\rho^\pm$ are long-lived with a decay length of order five centimeters.
  Here, the $\rho^0$ field is long-lived and neutral, giving rise to missing energy signatures.
\item Single charged-track and missing energy, Fig.~\ref{subfig:rhoMrhoZ}:
  one can have the associated production $pp \to \rho^0 \rho^- \to \rho^0 \rho^0 \pi^-$, and since the $\rho^-$ field
  is long lived one has a charged track until this field decays into $\rho^0$ and a pion.
  The $\rho^0$ field is neutral and long-lived. 
\item Multi-lepton channels, Figs.~\ref{subfig:psipsi2ee} and \ref{subfig:psipsi2mumu}:
  The pair production, $ p p \to \Psi^+ \Psi^- 
  \to \ell_i^{+} \ell_j^{-} \phi^* \phi$, give rise to two charged leptons and missing energy. 
  The $\phi$ field is neutral and long-lived in this context.
\end{itemize}

Note that the new fields predicted can be light, with masses close to the electroweak scale.
We study these signatures in detail to understand the possibility of testing this theory at the LHC.
In Ref.~\cite{Butterworth:2024eyr} we studied the experimental bounds on the leptophobic gauge boson, \ZB,
associated with the new local baryonic gauge symmetry, the missing energy signatures related to our dark matter
candidate, and the signatures from the new Higgs decays. In this paper we revisit \ZB production,
and discuss a new decay of the SM Higgs, $h \to \ZB \gamma$.
Our main findings tell us that a well-motivated theory for physics beyond the SM could be discovered in the
near future at the LHC.

The article is organized as follows:
In Sec.~\ref{theory} we discuss the minimal theory for spontaneous baryon number
and outline its main features.
In Section~\ref{sec:zb} revisit the phenomenology and limits on the \ZB,
with a note on the recently observed \ttbar excess. 
We examine the fermion decays, including some novel signatures, in Sec.~\ref{sec:fermions}.
In Sec.~\ref{Higgses} we return to the Higgs sector of the model, highlighting a new decay
for the SM-like Higgs.
We summarize our main findings in Sec.~\ref{Summary}.
In Appendix.~\ref{Rules} we provide the relevant Feynman rules, in Appendix~\ref{hZBgamma}
we show the explicit expression for the effective $h\ZB\gamma$ couplings.
\section{MINIMAL THEORY FOR BARYON NUMBER}
\label{theory}
In this theory, the Abelian global symmetry associated with baryon number in the SM is promoted to a local gauge
symmetry and the theory is based on the gauge
group~\cite{FileviezPerez:2010gw,FileviezPerez:2011pt,Duerr:2013dza,FileviezPerez:2014lnj,FileviezPerez:2024fzc}:
$$SU(3)_C \otimes SU(2)_L \otimes U(1)_Y \otimes U(1)_{B}.$$
This implies an additional gauge boson, \ZB, associated with the $U(1)_{B}$ symmetry. This boson must be given
mass by spontaneously breaking the new symmetry, leading to an additional CP-even Higgs boson, \HB.
In Ref.~\cite{FileviezPerez:2024fzc} it was pointed out that all the baryonic anomalies can be cancelled with only
four extra fermions, with the following quantum numbers:
\begin{eqnarray}
    \Psi_L & \sim & ({\bf{1}}, {\bf{1}},-1,3/4), \
    \Psi_R   \sim  ({\bf{1}},{\bf{1}},-1,-3/4), \nonumber \\
    \chi_L & \sim & ({\bf{1}},{\bf{1}},0,3/4), \ {\text{and}} \
    \rho_L   \sim  ({\bf{1}},{\bf{3}},0,-3/4). \nonumber
    \label{fermions}
\end{eqnarray}
The $\rho_L$ field can be written as
\begin{equation}
\rho_L=\frac{1}{\sqrt{2}} 
\left( \begin{matrix}
\rho_L^0 & \sqrt{2} \rho_L^+ \\
\sqrt{2} \rho_L^- & - \rho_L^0
\end{matrix}
\right).
\end{equation}
Mass can be generated for the new fermions by the same Higgs field that gives mass to the new gauge boson,
$S \sim ({\bf{1}},{\bf{1}},0,3/2)$, 
with the following Yukawa interactions:
\begin{eqnarray}
- \mathcal{L} &\supset& \lambda_\rho {\rm{Tr}}(\rho_L^T C \rho_L) S + \lambda_\Psi \bar{\Psi}_L \Psi_R S 
+ \lambda_\chi \chi_L^T C \chi_L S^* + \lambda_e \bar{\Psi}_L e_R \phi \ + \ {\rm{h.c.}}.
\label{interactions-masses}
\end{eqnarray}
The new scalar field, $\phi \sim ({\bf{1}},{\bf{1}},0,3/4)$, is introduced to allow the new electrically charged fields,
$\Psi_L$ and $\Psi_R$, to decay into a SM charged lepton and the neutral scalar
field, through the last interaction term in the above equation. 
There are thus three neutral fields, $\phi$, $\chi_L$ and $\rho^0_L$, present in the theory;
however, the heavier two of these can always decay to the lighter via higher-dimensional operators, meaning that there is only one dark matter
candidate in this context~\footnote{Notice that the dimension five operators: 
\begin{eqnarray}
- \mathcal{L} &\supset&   \frac{y_1}{\Lambda} \ell_L^T i \sigma_2 C \rho_L H \phi \ + \ 
 \frac{y_2}{\Lambda} \ell_L^T  i \sigma_2 C H \chi_L \phi^*  \ + \ \frac{y_3}{\Lambda} H^\dagger \chi_L^T C \rho_L H \ + \ {\rm{h.c.}},
 \label{hdops}
\end{eqnarray}
are allowed by the gauge symmetry.}.

The scalar potential in this theory is given by 
 \begin{eqnarray}
 V(H,S,\phi)&=&-m_H^2 H ^{\dagger}H+\lambda(H^{\dagger}H)^2-m_s^2 S ^{\dagger}S 
 + \lambda_s (S^{\dagger}S)^2-m_{\phi}^2 \phi ^{\dagger}\phi 
 + \lambda_{\phi}(\phi^{\dagger}\phi)^2 \nonumber \\
 &+& \lambda_1(H^{\dagger}H)S^{\dagger}S + \lambda_2(H^{\dagger}H)\phi^{\dagger}\phi 
 + \lambda_3(S^{\dagger}S)\phi^{\dagger}\phi + 
 \left( \mu S^* \phi \phi + {\rm{h.c.}} \right),
 \end{eqnarray}
and the scalar fields can be written as 
\begin{eqnarray}
H&=&\begin{pmatrix}
h^+\\
\frac{1}{\sqrt{2}}(v_{0} + h_0) e^{i \sigma_0/v_0} 
\end{pmatrix}, \
S = \frac{1}{\sqrt{2}}\left(v_{s} + h_s \right) e^{i \sigma_s/v_s}, 
\label{S-filed}
\textrm{and} \
\phi = \frac{1}{\sqrt{2}} \left( h_\phi + v_\phi \right)e^{i \sigma_\phi/v_\phi},
\end{eqnarray}
Here $h_i$ are the different CP-even Higgs fields and $v_i$ are the vacuum expectation values.

In general, there are two main scenarios, with very different predictions:
\begin{itemize}
\item $v_\phi=0$: In this case, if kinematically allowed, the new charged fermions can decay to the SM charged leptons and the field $\phi$.
  In this scenario, the lightest field among $\phi$, $\rho_L^0$, and $\chi_L$ can be a DM candidate because of the accidental
  discrete symmetry: 
  \begin{equation}
    {\mathcal{Z}}_2: \phi \to - \phi, \rho_L \to - \rho_L, \chi_L \to - \chi_L, \Psi_L \to - \Psi_L, \Psi_R \to - \Psi_R. 
  \end{equation}
Notice that this symmetry arises as a natural consequence of the spontaneous breaking of the gauge symmetry.
\item $v_\phi \neq 0$: In this case,
  all the new fields introduced to cancel the anomaly can decay to the SM fields.
  The field $\Psi_L$ can mix with the $e_R$, while the neutral fields, $\rho_L^0$ and $\chi_L$, decay via
  the higher-dimensional operators mentioned above. 
\end{itemize}
In the following, we focus on the first case, since in the second case, there is no DM candidate. This theory predicts the following physical states:
\begin{itemize}

\item \ZB is the gauge boson associated to the local baryon number.
  The \ZB mass is given by $\MZB=3 \gB v_S/2$ when 
  $v_\phi=0$.
\item $h$ is the SM-like Higgs boson defined as: $h= h_0 \ctB  - h_S \stB$.
\item \HB is the new CP-even Higgs defined as: $\HB=h_0 \stB + h_S \ctB$.
  In Ref.~\cite{Butterworth:2024eyr}, the production cross section of the new Higgs, and the experimental bounds
  on its mass, are discussed.
    
\item $\phi$ is a complex scalar field and it can be written as
  $\phi=(\phi_R+i\phi_I)/\sqrt{2}$. The masses for $\phi_R$ and $\phi_I$ read as: 
  \begin{eqnarray}
    M_{\phi_R}^2&=&-m_\phi^2+ \frac{\lambda_2}{2} v_0^2 + \frac{\lambda_3}{2} v_S^2 + \sqrt{2} \mu v_S, \\
    M_{\phi_I}^2&=& M_{\phi_R}^2 - 2 \sqrt{2} \mu v_S,
  \end{eqnarray}
  when $v_\phi=0$.
\item Two Majorana fermionic fields, 
  \begin{equation}
    \chi = \chi_L + (\chi_L)^C, {\rm{and}} \ \rho^0=\rho_L^0 + (\rho_L^0)^C,
  \end{equation}
  with masses given by
  \begin{eqnarray}
    M_\chi &=&\sqrt{2} \lambda_\chi v_S, \ {\text{and}} \
    M_{\rho^0} = {\sqrt{2}} \lambda_\rho v_S.
  \end{eqnarray}
\item Two charged fermionic fields: $\Psi^-$ and $\rho^-$ defined as
  \begin{eqnarray}
    \Psi^-&=&\Psi_L^- + \Psi_R^-, \ \rm{and} \
    \rho^- = \rho_L^- + (\rho_L^+)^C.
  \end{eqnarray}
\end{itemize}
For a detailed discussion of the dark-matter candidates in this context, see the study in Ref.~\cite{Debnath:2024cil}.
%
\section{THE LEPTOPHOBIC GAUGE BOSON}
\label{sec:zb}
A leptophobic gauge boson is predicted in the theory discussed in the previous section. The phenomenology and experimental bounds for the
specific \ZB gauge boson associated with this theory were discussed in detail in Ref.~\cite{Butterworth:2024eyr}.
This study showed that the mass of the \ZB can be close to the electroweak
scale and satisfy all collider bounds, without assuming a very small gauge coupling \gB.
We revisit this discussion briefly in light of new data, including new LHC measurements of \ttbar cross sections and a
reported excess close to threshold~\cite{CMS:2025kzt}.
Following the method described in Ref.~\cite{Butterworth:2024eyr}, we study \ZB production and decay using
\contur~\cite{Butterworth:2016sqg,Buckley:2021neu} 3.1 to compare predictions
to SM measurements from ATLAS and CMS. To do this, we have implemented our model in Feynrules~\cite{Alloul:2013bka} and exported a
UFO~\cite{Degrande:2011ua}
\footnote{Available from the \contur model library https://gitlab.com/hepcedar/contur/-/tree/main/data/Models}
directory which allows events to be generated by \herwig~7.3~\cite{Bewick:2023tfi,Bellm:2015jjp}
and passed through~\rivet~4.0.3~\cite{Bierlich:2024vqo,Buckley:2010ar}.

\begin{figure} [h]  
        \centering       
        \includegraphics[width=0.55\textwidth]{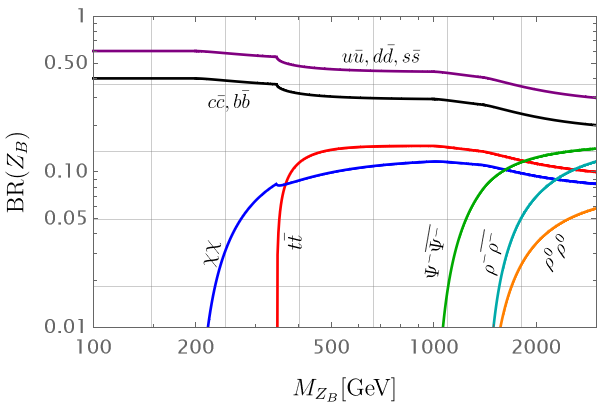}
        \caption{Branching ratios of the leptophobic gauge boson (\ZB) decays as a function of $M_{Z_B}$,
          assuming $M_{\rho^-}=700$ GeV and $M_{\Psi^-}=500$ GeV. These results are independent of the $g_B$ value.}
        \label{fig:ZBdecays}
\end{figure}

\begin{figure} [t]
      \centering
    \begin{subfigure}[t]{\textwidth}
    \centering
    \includegraphics[width=0.8\textwidth]{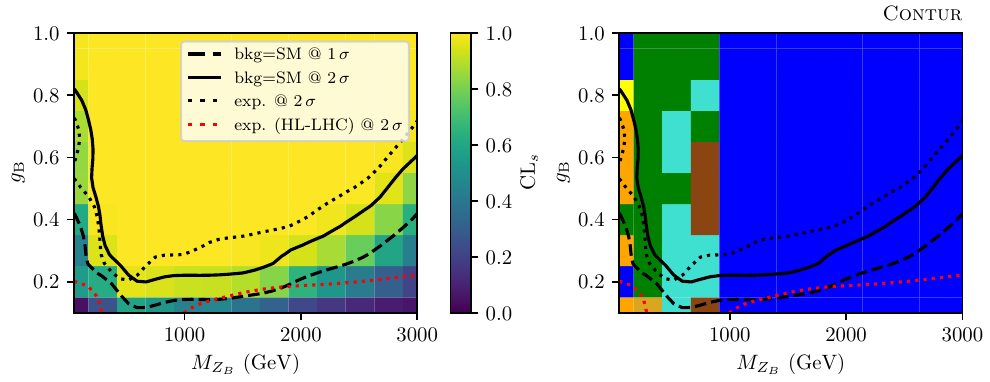} 
    \vspace{-0.5cm}
    \caption{}
    \end{subfigure}

    \begin{subfigure}[t]{\textwidth}
    \centering
    \includegraphics[width=0.8\textwidth]{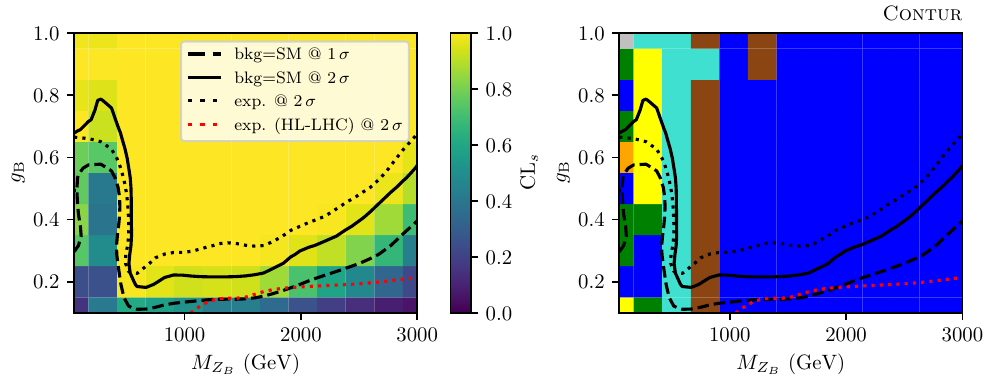}
    \vspace{-0.5cm}
    \caption{}
    \end{subfigure}

    \vspace{0.2cm} 
    
    \begin{subfigure}[t]{\textwidth}
        \centering

        \begin{tabular}{lll}
        \swatch{turquoise}~$\ell_1\ell_2$+\MET{}+jet \cite{ATLAS:2023gsl,ATLAS:2019ebv} & 
        \swatch{orange}~$\ell^+\ell^-$+jet \cite{ATLAS:2019ebv,ATLAS:2019ebv} & 
        \swatch{blue}~$\ell$+\MET{}+jet \cite{ATLAS:2017irc,CMS:2021vhb,ATLAS:2017luz,ATLAS:2015mip} \\ 
        \swatch{green}~\MET{}+jet \cite{ATLAS:2024vqf} &
        \swatch{saddlebrown}~hadronic $t\bar{t}$ \cite{ATLAS:2018orx,ATLAS:2020ccu} & 
        \swatch{goldenrod}~$\gamma$+\MET{} \cite{ATLAS:2018nci} \\
        \swatch{yellow}~$\gamma$ \cite{ATLAS:2017xqp,ATLAS:2019iaa} (in RH plots only)& 
        \swatch{silver}~jets \cite{ATLAS:2017ble} 
    \end{tabular}
    \end{subfigure}
    \caption{Exclusion limits in \gB as a functions of \MZB, (a) \Mchi = 100 GeV (b) \Mchi = 3 TeV. The solid black line indicates the 95\% exclusion and the dashed black line the 68\% exclusion. The dotted 
    black line is the expected exclusion, and the dotted red line is
    a naive estimate of the eventual HL-LHC sensitivity.}
    \label{fig:MZB_gB}
\end{figure}

In Fig.~\ref{fig:ZBdecays} we show the branching ratios of the leptophobic gauge boson decays as a function of $M_{Z_B}$,
assuming $M_{\rho^-}=700$ GeV and $M_{\Psi^-}=500$ GeV. Notice that $Z_B$ decays mainly to light quarks, but when the decays to new
fermions are kinematically allowed, they can have a large branching ratio. 
We repeat the study of Fig.~5 from Ref.~\cite{Butterworth:2024eyr}, which investigated the sensitivity of the LHC data
present in \contur to the \ZB as function of its mass, \MZB, in the presence of a DM candidate $\chi$ with a mass of 100~GeV.
We also present the case in which the $\chi$ decay is not kinematically accessible. We show the results in Fig.~\ref{fig:MZB_gB}. There are several changes with respect to
\cite{Butterworth:2024eyr}: further measurements have been added to \contur, notably of leptonically-decaying top cross sections;
the SM uncertainties for the hadronic \ttbar and $Z$+jets are now included; and the model studied here differs
in the $\ZB \rightarrow \chi\chi$ branching ratio by a factor of four because the $\chi_L$ field has a smaller baryon number in
the minimal model we discuss in this article.
However, the results are qualitatively very similar, with $\gB < 0.25$ allowed for $\MZB \approx 600$~GeV, with higher values
allowed at higher and lower \MZB values.
Close to the \ttbar threshold, the upper limit on \gB is around 0.25. It is important to mention that the most sensitive channels are the hadronic \ttbar and the $\ell + \MET{}+{\rm jet}$, when the $Z_B$ does not decay to the dark-matter candidate; see details in the right-panel in Fig.~\ref{fig:MZB_gB}.

\begin{figure} [t]  
        \centering       
        \includegraphics[width=0.55\textwidth]{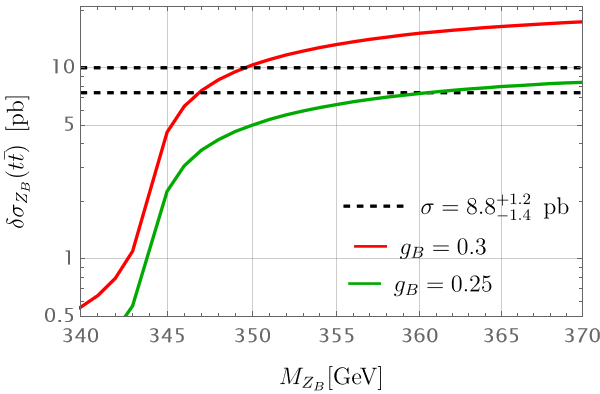}
        \caption{Extra contribution to the  production cross section of $t \bar{t}$ due to $Z_B$ at $\sqrt{s}=13$ TeV, as a function of \MZB,
          for two different \gB values,
          calculated using \madgraph ~\cite{Alwall:2014hca}.
          The cross section excess reported by CMS is also indicated by the dashed black lines.}
        \label{fig:ttx}
\end{figure}

The enhancement to the \ttbar cross section due to the \ZB, as function of \MZB, is shown in Fig.~\ref{fig:ttx}. Near the \ttbar threshold, a cross section of several  fb can be
accommodated: this is confirmed by a fit using \herwig  with \contur\footnote{Making use of the \spey~\cite{Araz:2023bwx}
functionality introduced in version 3.1.} which
gives a maximum cross section for $pp \rightarrow \ZB \rightarrow \ttbar$ of 9.2~pb at 95\% c.l., for \MZB = 360~GeV.
The CMS collaboration have recently reported an excess in this region consistent with the production of
a pseudoscalar \ttbar bound state with a production cross section of $8.8^{+1.2}_{-1.4}$~pb, while stating that other explanations
cannot be ruled out~\cite{CMS:2025kzt}. We note that one such explanation might be a \ZB with mass around 360~GeV, although
without a detailed study including the experimental acceptance for a vector state, nothing more definite can be said at this stage.
%
\section{FERMION DECAYS}
\label{sec:fermions}

The theory predicts two charged fermionic fields $\Psi^-$ and $\rho^-$ with masses given by
\begin{eqnarray}
M_{\Psi^-} &=& \frac{1}{\sqrt{2}} \lambda_\Psi v_S, \ {\text{and}} \
M_{\rho^-}  =  M_{\rho^0} + \Delta M,    
\end{eqnarray}
where $M_{\rho^0}=\sqrt{2} \lambda_\rho v_S$ and the mass splitting $\Delta M \approx 166$ MeV is generated at one-loop level~\cite{Cirelli:2005uq,Belyaev:2022qnf}.
Since this mass splitting is very small, the dominant allowed decays are $\rho^- \to \rho^0 \pi^-, \rho^0 e^- \bar{\nu}$ and
$\rho^0 \mu^- \bar{\nu}$, with widths given by
\begin{eqnarray}
    \Gamma (\rho^- \to \rho^0 \pi^-) &=& \frac{2 G_F^2 V_{ud}^2 \ \Delta M^3 f_\pi^2}{\pi} \sqrt{1 - M_\pi^2/\Delta M^2},\\
    \Gamma (\rho^- \to \rho^0 e^- \bar{\nu})&=& \frac{ 2 G_F^2  \Delta M^5}{15 \pi^3},\\
    \Gamma (\rho^- \to \rho^0 \mu^- \bar{\nu})&=& 0.12 \ \Gamma (\rho^- \to \rho^0 e^- \bar{\nu}).
\end{eqnarray}
Here, $f_\pi=131$ MeV, is the decay constant of the pion.
The $\rho^-$ decays mainly into a pion and $\rho^0$ with a branching ratio
$\rm{BR}(\rho^- \to \rho^0 \pi^-) \sim 97 \%$. The decay length is given by
 \begin{equation}
     c \tau_{\rho^-} \approx 5.6 \ \textrm{cm}.
 \end{equation}
 Therefore, the charged $\rho^-$ is a long-lived charged particle, which will give rise to tracks in a charged-particle detector which terminate in a `kink' at the decay point, from where a soft pion track will emerge. The general topology is
 illustrated in Figs.~\ref{subfig:rhoMrhoMbar} and \ref{subfig:rhoMrhoZ}. Notice that this prediction could change if the higher-dimensional operators in Eq.(\ref{hdops}) are allowed. In this case, $\rho^-$ can decay into a charged lepton and the scalar field $\phi$.
 
\begin{figure} [t]  
        \centering       
        \includegraphics[width=0.6\textwidth]{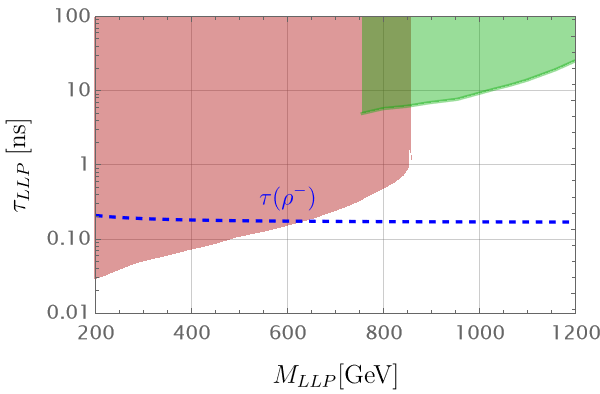}
        \caption{Allowed parameter space for a charged long-lived particle (LLP). The red shaded region is excluded by the
          ATLAS measurements~\cite{ATLAS:2022rme}, and the green shaded region is excluded by the studies in Ref.~\cite {atlasLLP}.
          The dashed line shows the prediction of this model.}
        \label{fig:LLP}
\end{figure}

 From the ATLAS and CMS searches looking for a long-lived charged pure wino field produced through the weak interactions,
 one can conclude that the mass of the $\rho^-$ field has to be above 650~GeV, see Refs.~\cite{ATLAS:2022rme,CMS:2023mny}
 for details. The experimental bounds estimated from reinterpreting these searches are shown in Fig.~\ref{fig:LLP}.
 For our theory, masses above 650~GeV are still allowed.
 These channels with charged tracks are quite distinctive and can be used to test this theory in the near future.
 See also Ref.~\cite{Giudice:2022bpq} for a recent discussion about charged tracks in new physics models, and
 Ref.~\cite{ATLAS:2025fdm} for recent experimental bounds.

\begin{figure}[t]
  \centering
  \begin{subfigure}[b]{0.45\textwidth}
    \centering
    \includegraphics[width=\textwidth]{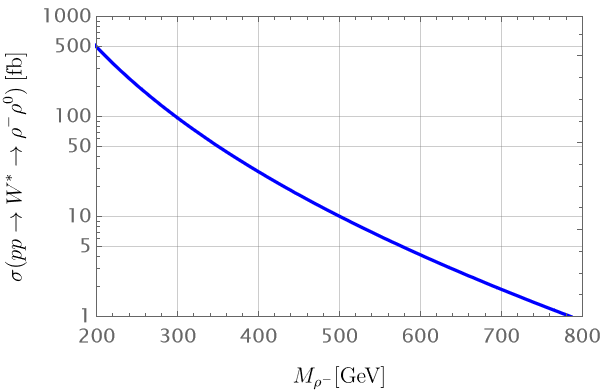}
    \caption{}
    \label{subfig:DY_rhoMrhoZ}
  \end{subfigure} 
  \begin{subfigure}[b]{0.45\textwidth}
    \centering
    \includegraphics[width=\textwidth]{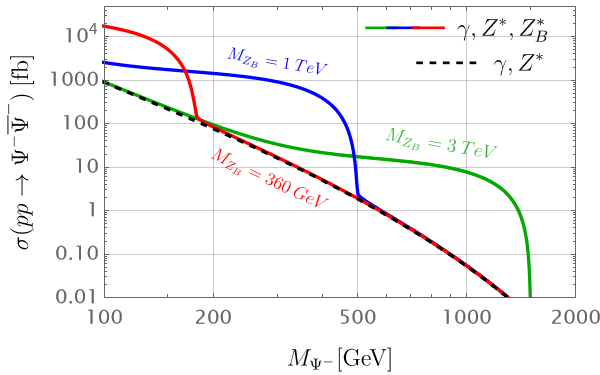}
    \caption{}
    \label{subfig:DY_psipsi}
  \end{subfigure}  
  \caption{Leading order Drell-Yan production cross sections of the new fermions at $\sqrt{s}=13$ TeV,
    as a function of the charged fermion mass, calculated using \madgraph~\cite{Alwall:2014hca}.
    (a) $\rho^0 \rho^-$ 
    (b) $\Psibar {\Psi}^-$, production at the LHC with $\gB=0.25$ and three different \ZB masses, as well
    as the prediction when the \ZB plays no role.}
    \label{fig:prodcross}
\end{figure}

Fig.~\ref{subfig:DY_rhoMrhoZ} shows the $\rho^0 \rho^-$ production cross section at $\sqrt{s}=13$~TeV as a function
 of the charged fermion mass, calculated using \madgraph~\cite{Alwall:2014hca}. In Fig.~\ref{subfig:DY_psipsi} we show the
predicted cross section for $\Psibar\Psi^-$ production via the SM $ \gamma ,Z$ and the \ZB, for three different
\ZB masses, as well as for the case in the which \ZB plays no role. As can be seen, the \ZB contribution can significantly
change the cross section around the $2 \MPsi \sim \MZB$ threshold.

The $\Psi^-$ can decay into a charged lepton and the scalar field $\phi$, if this channel is kinematically allowed.
In the opposite case, $M_{\Psi^-} < M_\phi$, the $\Psi^-$ field decays via higher-dimensional operators and it is generically long-lived giving rise to a charged track.
The $\Psi^-$ decays are represented schematically in Fig.~\ref{subfig:psipsi2ee} and \ref{subfig:psipsi2mumu}. 

\begin{figure}[t]
    \centering
    \begin{subfigure}[t]{\linewidth}
    \centering
        \includegraphics[width=0.8\textwidth]{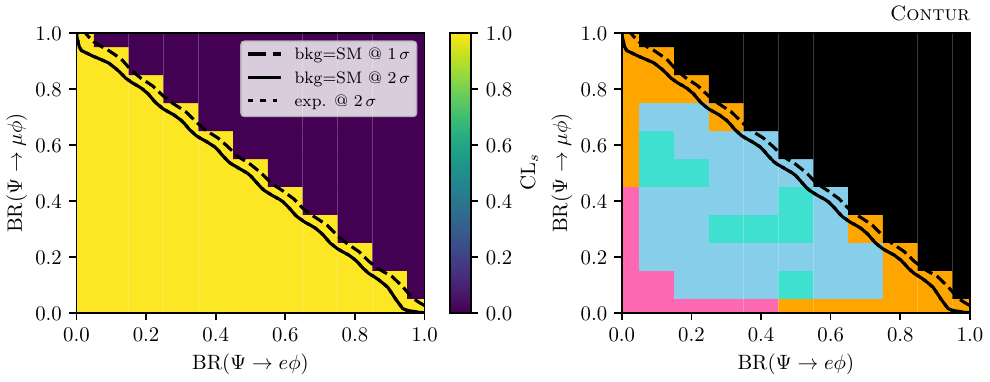}
    \end{subfigure}


    \begin{subfigure}[t]{\textwidth}
        \centering

        \begin{tabular}{llll}
            \swatch{orange}~$\ell^+\ell^-$+jet \cite{CMS:2022ubq} &
            \swatch{skyblue}~$\ell_1\ell_2$+\MET{} \cite{ATLAS:2019rob} & 
            \swatch{turquoise}~$\ell_1\ell_2$+\MET{}+jet \cite{ATLAS:2021jgw} & 
            \swatch{hotpink}~$\tau^+\tau^-$ \cite{ATLAS:2025oiy}
        \end{tabular}
    \end{subfigure}
    \caption{(Left) Exclusion for each point in the grid of branching ratios, with interpolated excluded contours overlaid.
      (Right) The measurement with the highest exclusion power at each point.
      In this scenario $\MPsi = 300$~\GeV, $\Mphi=100$~\GeV, $\MZB=1$~\TeV \ and \gB = 0.25.}
    \label{fig:BR_triangle_mpsi_300}
\end{figure}
\begin{figure}[t]
  \centering
    \begin{subfigure}[c]{0.45\textwidth}
        \includegraphics[width=\linewidth]{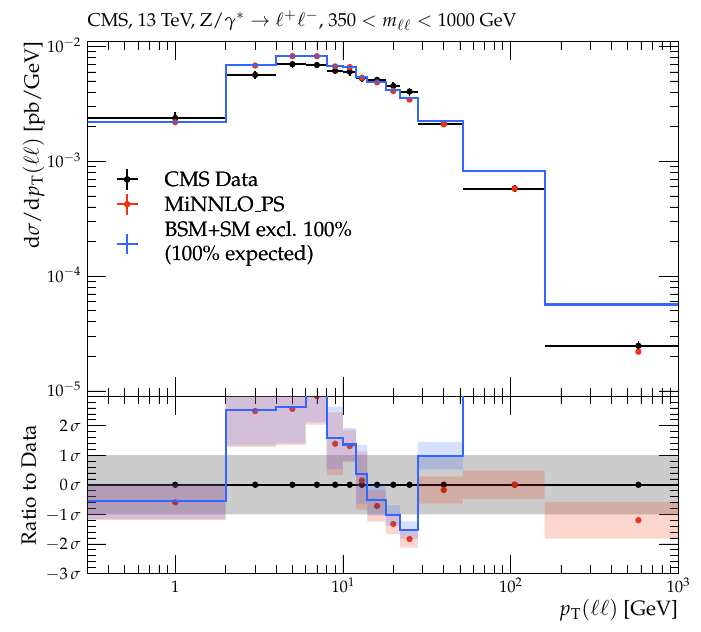}
         \caption{}
         \label{subfig:rivetplots_a}
    \end{subfigure}
    \begin{subfigure}[c]{0.45\textwidth}
        \includegraphics[width=\linewidth]{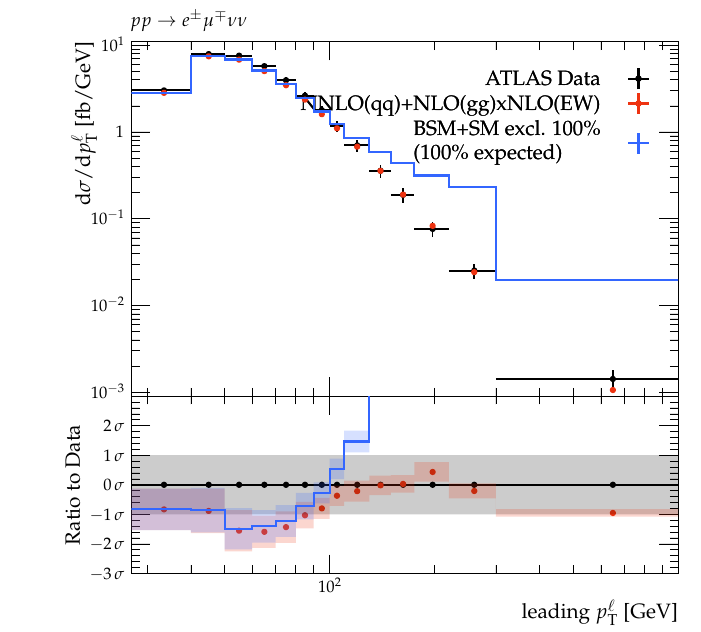}
        \caption{}
        \label{subfig:rivetplots_b}
    \end{subfigure}    
    \begin{subfigure}[c]{0.45\textwidth}
        \includegraphics[width=\linewidth]{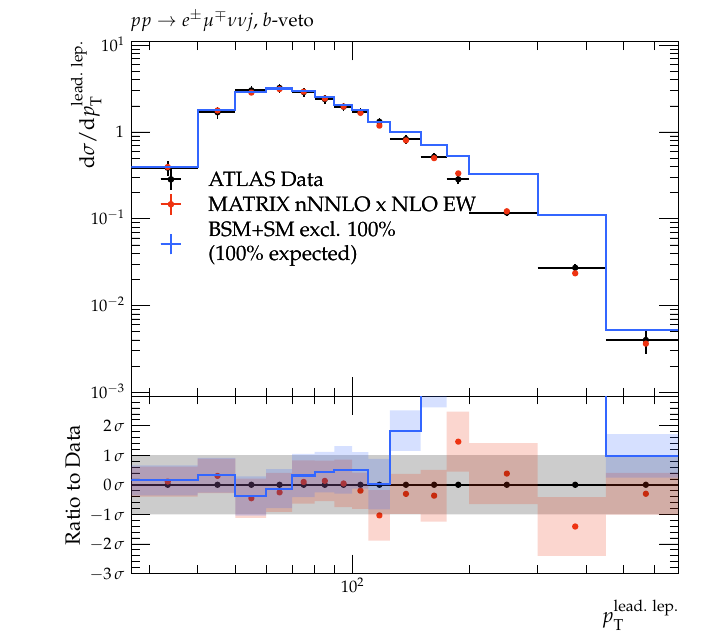}
        \caption{}
        \label{subfig:rivetplots_c}
    \end{subfigure}    
    \begin{subfigure}[c]{0.5\textwidth}
        \includegraphics[width=\linewidth]{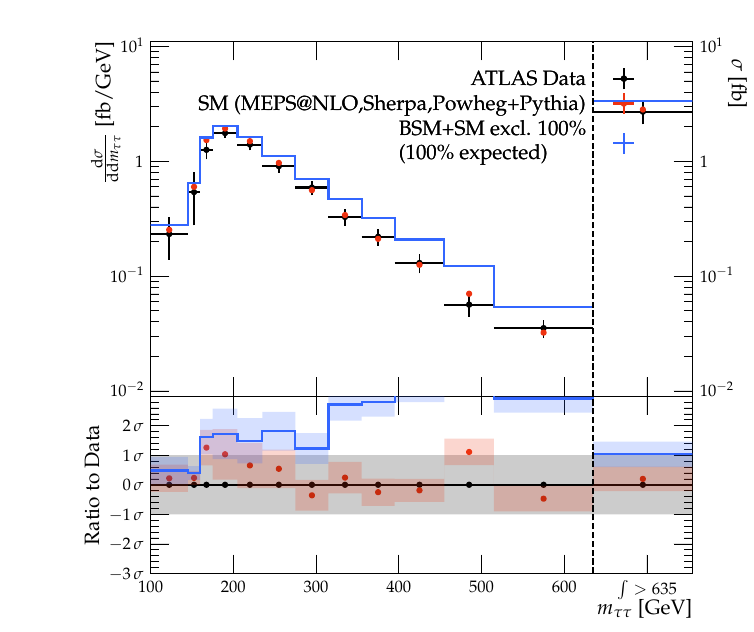}
        \caption{}
        \label{subfig:rivetplots_d}
         \label{fig:ATLAS_ditau}     
    \end{subfigure}
    \caption{BSM+SM vs SM only hypotheses for various distributions providing exclusion at parameter points from
      Fig.~\ref{fig:BR_triangle_mpsi_300}.
      (a) Dilepton transverse momentum distribution from~\cite{CMS:2022ubq,Monni:2020nks,Monni:2019whf}.
      The signal hypothesis is excluded by an excess of events in the high momentum tail.
      At this point in the grid, $\text{BR}(\Psi \rightarrow e \phi)$~=~1.
      (b) Leading lepton transverse momentum distributions from Ref.~\cite{ATLAS:2019rob,Cascioli:2011va,Grazzini:2017mhc,Caola:2016trd,Biedermann:2016guo,Gehrmann:2015ora} and (c) Ref.~\cite{ATLAS:2021jgw,Grazzini:2017mhc,Grazzini:2016ctr,Kallweit:2014xda,Gehrmann:2014fva,Cascioli:2011va}.
      At this point in the grid, $\text{BR}(\Psi \rightarrow e \phi)=\text{BR}(\Psi \rightarrow \mu \phi)$~=~0.5.
      (d) Visible $\tau^+\tau^-$ invariant mass distribution~\cite{ATLAS:2025oiy,Sherpa:2019gpd}.
      At this point in the grid, $\text{BR}(\Psi \rightarrow \tau \phi)$~=~1.
      }
    \label{fig:rivetplots}     
\end{figure}
Since the production and decay of $\Psi^-$ leads to SM-like signatures involving charged leptons and missing transverse momentum,
we may also study these using \contur.
As an illustrative example, we assume \MPsi=300~GeV, \Mphi=100~GeV, \MZB=1~TeV and \gB=0.25. The branching ratio of $\Psi$ to each of $e$, $\mu$ and $\tau$ varies over the
full range between zero and unity. The results are shown in Fig.~\ref{fig:BR_triangle_mpsi_300}.
Points above the grid diagonal are unphysical, since the sum of the branching fractions exceeds unity.

In the top-left and bottom-right regions, the $\Psi$ decays with a high branching fraction to electrons or muons, respectively.
This results in a same-flavour lepton pair in the final state, which is constrained primarily by the CMS Run~2 high-mass
Drell-Yan measurement~\cite{CMS:2022ubq}, as can be seen in the right-hand panel of the figure,
which indicates the measurements providing the strongest sensivity at each parameter point.
An example differential distribution with high exclusion power is shown in Fig.~\ref{subfig:rivetplots_a}.
In the central region of the grid in Fig.~\ref{fig:BR_triangle_mpsi_300}, the branching fraction to electrons and muons
is approximately equal, leading to more final states with electrons and/or muons and missing energy.
This scenario is constrained by measurements of $WW$ production in the unlike dilepton final
state~\cite{ATLAS:2019rob,ATLAS:2021jgw}, as shown by right panel.
Example differential distributions in this region are shown in Figs.~\ref{subfig:rivetplots_b} and \ref{subfig:rivetplots_c}.

\begin{figure}[tbph]
    \centering
    \begin{subfigure}[t]{\linewidth}
    \centering
        \includegraphics[width=0.8\textwidth]{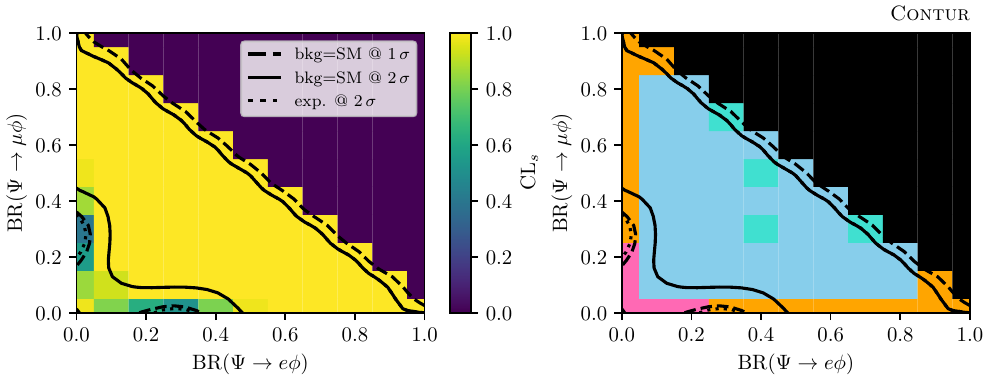}
    \end{subfigure}


    \begin{subfigure}[t]{\textwidth}
        \centering

        \begin{tabular}{llll}
            \swatch{orange}~$\ell^+\ell^-$+jet \cite{CMS:2022ubq} &
            \swatch{skyblue}~$\ell_1\ell_2$+\MET{} \cite{ATLAS:2019rob} & 
            \swatch{turquoise}~$\ell_1\ell_2$+\MET{}+jet \cite{ATLAS:2021jgw} & 
            \swatch{hotpink}~$\tau^+\tau^-$ \cite{ATLAS:2025oiy}
        \end{tabular}
    \end{subfigure}

    \caption{Same as Figure~\ref{fig:BR_triangle_mpsi_300} but with $\MPsi = 400$~\GeV.}
    \label{fig:BR_triangle_mpsi_400}
\end{figure}

\begin{figure}[tbp]
    \centering
    \begin{subfigure}[t]{\linewidth}
    \centering
    \includegraphics[width=0.8\textwidth]{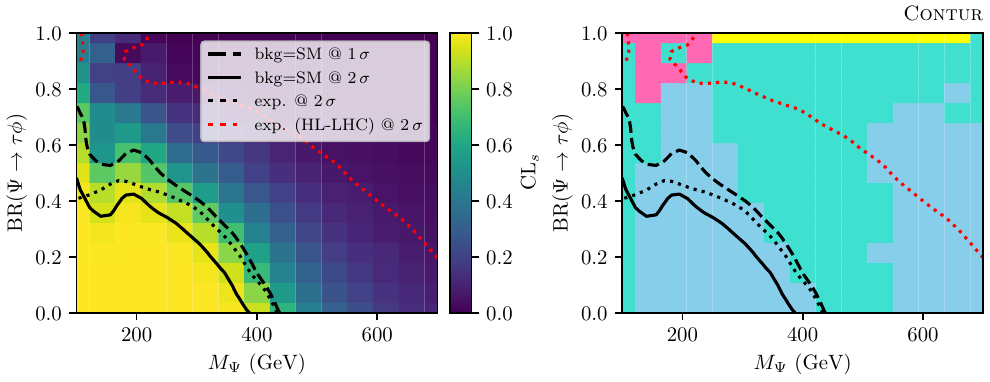}
    \vspace{-0.5cm}
    \caption{}
    \label{subfig:no_zb}
    \end{subfigure}

    \begin{subfigure}[t]{\linewidth}
    \centering
    \includegraphics[width=0.8\textwidth]{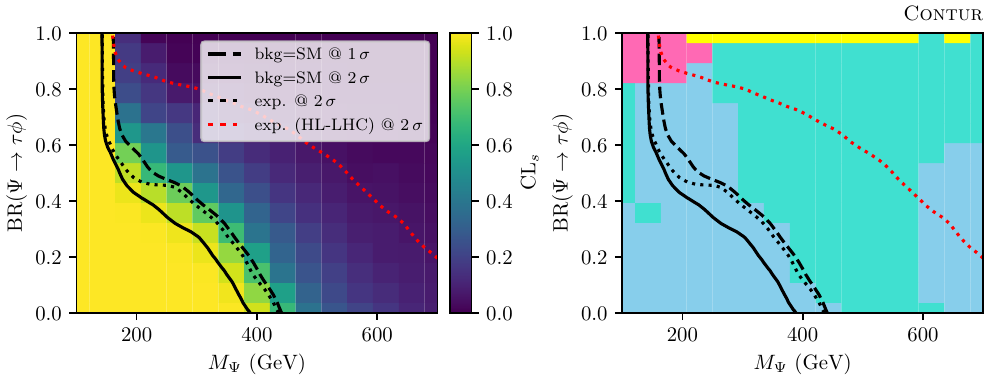}
    \vspace{-0.5cm}
    \caption{}
    \label{subfig:mzb_360GeV}
    \end{subfigure}

    \begin{subfigure}[t]{\linewidth}
    \centering
    \includegraphics[width=0.8\textwidth]{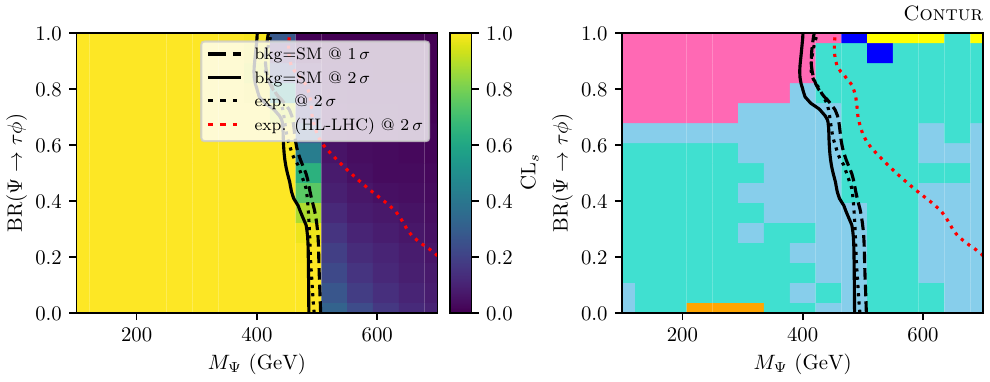}
    \vspace{-0.5cm}
    \caption{}
    \label{subfig:mzb_1TeV}
    \end{subfigure}

    \begin{subfigure}[t]{\linewidth}
    \centering
    \includegraphics[width=0.8\textwidth]{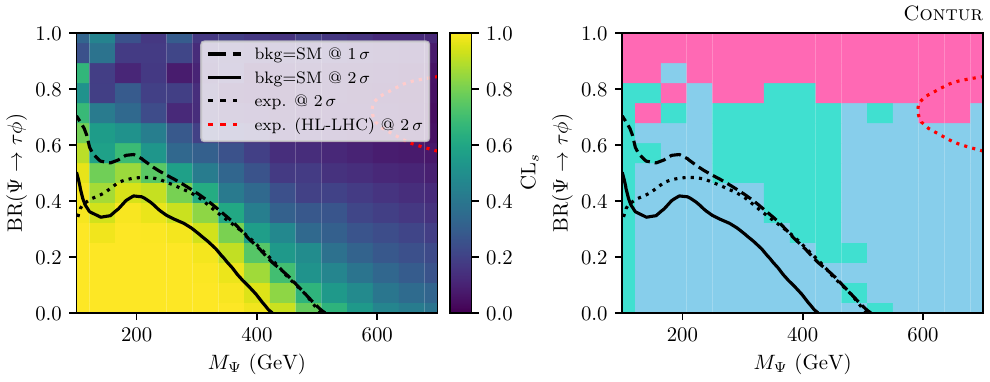}
    \vspace{-0.5cm}
    \caption{}
    \label{subfig:mzb_3TeV}
    \end{subfigure}

    \begin{subfigure}[t]{0.9\linewidth}
        \centering
        \begin{tabular}{llll}
        \swatch{blue}~$\ell$+\MET{}+jet \cite{ATLAS:2017irc,CMS:2021vhb,ATLAS:2017luz,ATLAS:2015mip} & 
        \swatch{skyblue}~$\ell_1\ell_2$+\MET{} \cite{ATLAS:2019rob} & 
        \swatch{turquoise}~$\ell_1\ell_2$+\MET{}+jet \cite{ATLAS:2021jgw} & 
        \swatch{orange}~$\ell^+\ell^-$+jet \cite{CMS:2022ubq} \\ 
        \swatch{yellow}~$\gamma$ \cite{ATLAS:2017xqp,ATLAS:2019iaa} & 
        \swatch{hotpink}~$\tau^+\tau^-$ \cite{ATLAS:2025oiy} & 
    \end{tabular}
    \end{subfigure}

    \caption{Exclusion contour in the plane of the $\Psi$ mass and its branching ratio to third generation leptons,
      assuming branching to the 1st and 2nd generations are equal.
      The mass of $\phi$ was taken to be 50~\GeV and $\Mchi = 500$~GeV.
      In (a) $\gB=10^{-6}$, removing the \ZB contribution to $\Psi^+\Psi^-$ production.
      In the other plots, $\gB=0.25$ and \MZB is (b) 360~GeV, (c) 1~TeV and (d) 3~TeV.}
    \label{fig:psi_mass_BR}
\end{figure}

In the bottom-left region of Fig.~\ref{fig:BR_triangle_mpsi_300}, the branching ratio is primarily to $\tau$-leptons, which
is more challenging experimentally. Nevertheless, the right panel of Fig.~\ref{fig:BR_triangle_mpsi_300} shows that this
region of parameter space is excluded by the recent ATLAS di-$\tau$ measurement~\cite{ATLAS:2025oiy}.
The distribution driving this exclusion is shown in Fig.~\ref{subfig:rivetplots_d}.
As expected, the sensitivity drops as the mass of the $\Psi$ increases. 
Fig.~\ref{fig:BR_triangle_mpsi_400} shows the same scenario as Fig.~\ref{fig:BR_triangle_mpsi_300}, but with the $\Psi$ mass increased to 400~GeV.
In this case the scenarios involving decays to muons and/or electrons are still excluded, but the more
challenging di-$\tau$ final states are less constrained.
Since the muon and electron sensitivities are quite similar, we set them equal to
each other, which allows us to perform a scan over the mass of the $\Psi$ and the branching ratio
to $\tau$-leptons, with the remainder of the decays being equally to electrons or muons. This is shown
in Fig.~\ref{fig:psi_mass_BR}, for various \ZB mass scenarios, since, as illustrated in Fig.\ref{fig:prodcross}, the size of
the contribution will depend upon \MZB and \gB. We consider four scenarios; one with
the \ZB decoupled; one motivated by the possibility of a resonant contribution at top threshold ($\MZB = 360 \ \GeV, \gB = 0.25$)
as discussed in Section~\ref{sec:zb}; one where the \ZB is more massive but within reach ($\MZB = 1 \ \TeV, \gB = 0.25$),
and one where is it
more massive still, and thus contributes primarily as an off-shell propagator ($\MZB = 3 \ \TeV, \gB = 0.25$).

When the \ZB plays no role (Fig.~\ref{subfig:no_zb}) the model is ruled out for $\MPsi < 350$~GeV when the
$\Psi$ decays mostly to $\mu$ or $e$, but there is currently no significant exclusion when $\tau$ decays dominate.
However, a naive\footnote{This is likely to underestimate
the eventual HL-LHC reach, as discussed for example in Ref.~\cite{Belvedere:2024wzg}.}
estimate of HL-LHC sensitivity, based upon scaling the experimental uncertainties with the
square root of the ratio of the current integrated luminosity to 3~ab$^{-1}$,
shows that a $\Psi$ decaying predominantly to $\tau$-leptons could
be within reach for $\MPsi$ up to at least 200~GeV.
For a \ZB coupled with $\gB = 0.25$, and $\MZB = 360$~GeV, (Fig.~\ref{subfig:mzb_360GeV}), the situation is similar
except that there is some current exclusion up to $\MPsi \approx 140$~GeV.
For a \ZB coupled with $\gB = 0.25$, and $\MZB = 1$~TeV, (Fig.~\ref{subfig:mzb_1TeV}), the model
is excluded for $\MPsi < 400$~GeV when $\Psi \rightarrow \tau\tau$ dominates, and for $\MPsi < 480$~GeV otherwise.
Finally, for a \ZB coupled with $\gB = 0.25$ but $\MZB = 3$~TeV, (Fig.~\ref{subfig:mzb_3TeV}), the situation
is similar to Fig.~\ref{fig:psi_mass_BR}a with the \ZB decoupled, except that the expectation is that deviations
in the tails of the lepton distribution are expected to be sensitive for HL-LHC data, regardless of whether the $\Psi$ decays to
$e$, $\mu$ or $\tau$.
Note that in these results, only the sensitivity from $\Psi^+ \Psi^-$ production is evaluated; the production of \ZB itself
would provide additonal sensitivity in other final states, as already discussed in Section~\ref{sec:zb}.

The main conclusion is that while some of the parameter space is already ruled out by LHC measurements,
a large and well-motivated region remains, much of which is nevertheless in reach of HL-LHC data.
Therefore, one can hope to test these predictions in the near future.

\section{HIGGS DECAYS}
\label{Higgses}
Finally, we return to the Higgs sector and consider the decays of the \HB and the SM-like Higgs in this theory.
\subsection{Cucuyo Higgs Decays}
\begin{figure} [h]  
        \centering       
        \includegraphics[width=0.65\textwidth]{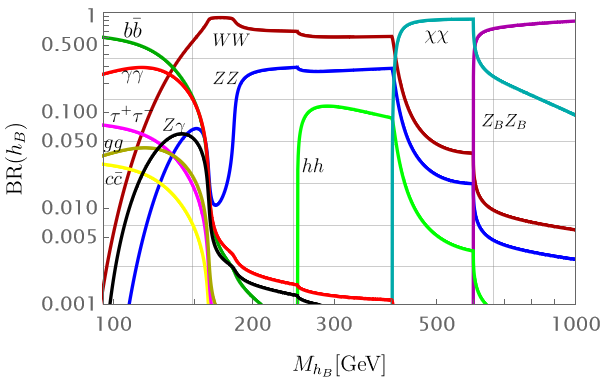}
         \caption{Branching ratios for \HB decays. Here we use $\gB = 0.25$, $\stB=0.01$, $M_\chi=200$ GeV, $\MZB=300$ GeV, $M_{\Psi^-}=1$ TeV, and $M_{\rho^-}=800$ GeV.}
         \label{Cucuyo}
        
\end{figure}

The new physical Higgs boson in the theory, \HB, can decay into the SM fields and the new fields. The main decays are:
\begin{equation}
\HB \to \bar{b}b, \gamma \gamma, WW, ZZ, \chi \chi, \ZB \ZB.    
\end{equation}
We refer to this new Higgs as ``Cucuyo Higgs'' following the discussion in Ref.~\cite{Butterworth:2024eyr}.
In Fig.~\ref{Cucuyo} we show the branching ratios for \HB decays.
Here we use $\gB = 0.25$, $\stB=0.01$, $M_\chi$=200 GeV, $\MZB=300$ GeV, $\MPsi=1$ TeV, and $M_{\rho^-}=800$ GeV.
Similar results were presented in Ref.~\cite{Butterworth:2024eyr} but in the current case the fields needed for anomaly
cancellations are different, which has implications for the predicted decays.
Fig.~\ref{Cucuyo} shows that, as in Ref.~\cite{Butterworth:2024eyr}, the \HB can have a large branching ratio into two photons when its mass is below
the $WW$ threshold. It can decay mainly into the dark matter candidate if the relevant masses allow, and it decays mainly into the new gauge boson
when this decay is kinematically allowed.
The production cross section and the experimental bounds on the Cucuyo Higgs mass were discussed in detail in Ref.~\cite{Butterworth:2024eyr} and
will still apply here with modifications given by the new branching ratios.

\subsection{Decay of the SM-like Higgs}
In this theory the SM-like Higgs can have a new decay at one-loop level due to the fact that the new gauge boson, \ZB,
can have $\MZB < 125$~\GeV. 
The effective $h\ZB\gamma$ coupling is generated at one-loop level,
where inside the loop one has the top quark.
The decay width for this process is given by
\begin{equation}
\Gamma (h \to \gamma \ZB)=  
\frac{ \alpha \ \alpha_B M_t^2 Q_t^2 |A|^2 \left( M_h^2-\MZB^2\right)}{128 \ \pi^3 \ v_0^2 \ M_h^3}.
\end{equation}
Here $\alpha=e^2/4 \pi$, $Q_t=2/3$ and $\alpha_B=\gB^2/4 \pi$. The explicit form of the $A$ coefficient is given in Appendix~\ref{hZBgamma}.
In Fig.~\ref{Brh} we show the branching ratio for $h \to \gamma \ZB$ for different values of the gauge coupling \gB and new gauge boson mass $\MZB$,
including the recent LHC bounds taken from~\cite{CMS:2024ztr}.
%
\begin{figure} [t]  
        \centering       
        \includegraphics[width=0.65\textwidth]{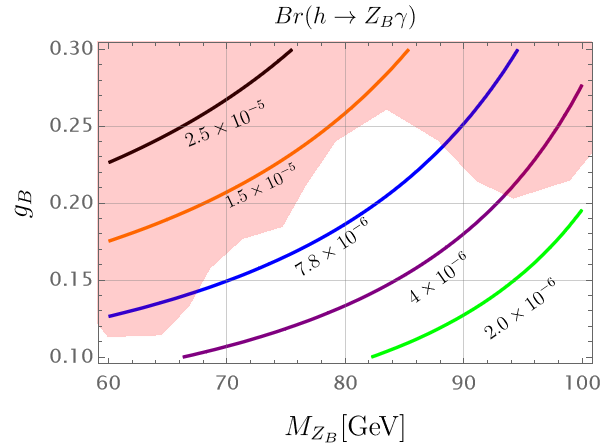}
        \caption{Branching ratio for $h \to \ZB \gamma $. The red region is excluded by LHC searches~\cite{CMS:2024ztr}.}
         \label{Brh}
\end{figure}
Unfortunately, the branching ratio for this decay is very small. Using the gluon fusion production cross section for the SM Higgs at
$\sqrt{s}=13$ TeV, $\sigma (pp \to h)=48.6$ pb~\cite{ParticleDataGroup:2024cfk}, we make a naïve estimate of the number of events for
the $\ZB \gamma$ channel:
\begin{equation}
    \sigma (pp \to h) \times {\rm{BR}}(h \to \ZB \gamma) \times 3000 \ {\rm{fb^{-1}}} \sim 146,
\end{equation}
using ${\rm{BR}}(h \to \ZB \gamma) \sim 10^{-6}$ as an example.
At the LHC, large QCD backgrounds make this decay very challenging to identify,
although given the significant improvements being delivered for example in jet tagging using new QCD insights coupled with machine learning,
see for example Ref.~\cite{Belvedere:2024wzg,ATLAS:2023krw,CMS:2021zto}, we hesitate to completely dismiss the possibility of observation.
At future colliders, a dedicated measurement of this channel becomes more feasible.
\section{SUMMARY}
\label{Summary}
The distinctive predictions of the minimal theory for local baryon number,
in which the global symmetry
associated with baryon number in the SM is promoted to a local gauge symmetry, have been explored.
The consequences of the spontaneous breaking of this symmetry were studied, identifying several novel phenomena.
The minimal framework requires only four additional fermionic representations to ensure anomaly cancellation, plus an additional scalar field.
Notably, it predicts the existence of a leptophobic gauge boson, \ZB, a new ``Cucuyo'' Higgs boson, \HB with a potentially large branching
ratio into photons, and new fermionic states with intriguing decay patterns.
In this theory, the SM Higgs can decay into a photon and \ZB when kinematically allowed, although the branching ratio is small,
making this decay very difficult to observe at the LHC.
The \HB decays were revisited, showing the branching ratios and computing the decay into two photons due to the presence of the
new charged fermions needed for anomaly cancellation. 

The long-lived fermion field, $\rho^-$, can be produced either in pairs or alongside its neutral partner, leading to distinct signatures
featuring ``kinked'' tracks and missing momentum.
Additionally, multi-lepton events and missing energy signatures offer intriguing prospects.
We established a lower bound on the $\rho^-$ mass by analyzing searches for charged tracks at the LHC, while for $\Psi^-$,
we identified the allowed parameter space based on various leptonic decay channels.
Di-lepton measurements, including those of $\tau$-leptons, play an important role in constraining the parameter space.
Within the allowed pararmeter space, it is possible that $\ZB \rightarrow \ttbar$ decays could play a role in explaining the excess
reported recently by CMS in the \ttbar cross section near threshold.

These findings suggest that a well-motivated and unique gauge theory beyond the Standard Model could describe physics below the TeV scale while
remaining consistent with all experimental constraints. The predictions of the theory could be observed in data to be collected
over the next years at the HL-LHC.

{\textit{Acknowledgements}}: P.F.P. thanks the SIMONS Foundation for financial support during his stay at the Galileo Galilei
Institute for Theoretical Physics in Florence, Italy. J.C.E. is supported by the STFC UCL Centre for Doctoral Training in Data Intensive Science (grant ST/W00674X/1) including departmental and industry contributions.
This work made use of the High Performance Computing Resource in the Core Facility
for Advanced Research Computing at Case Western Reserve University.
\appendix
\section{FEYNMAN RULES}
\label{Rules}
\begin{align}
    q \bar{q} \ZB  &: \hspace{0.5cm}  i\frac{\gB}{3} \gamma^\mu, \\
    \chi \chi \ZB &: - i \frac{3}{4} \gB \gamma^\mu \gamma^5, \\
    {\overline{\rho^-} \rho^- \ZB} &: {i \frac{3}{4} \gB \gamma^\mu \gamma^5},\\
      \overline{\Psi^-} \Psi^- \ZB  &: -i \frac{3}{4} \gB  \gamma^\mu \gamma^5, \\
     \chi \chi \HB &: i \frac{3}{2} \frac{\gB M_\chi}{\MZB} \ctB, \\
     \ZB \ZB \HB &: - i {3} \gB \MZB \ctB \ g^{\mu \nu}, \\
      Z Z \HB & : \hspace{0.5 cm}  2i \frac{M_{Z}^2}{v_0} \sin{\tB} \ g^{\mu \nu}, \\
      W W \HB & : \hspace{0.5 cm}  2i   \frac{M_{W}^2}{v_0} \sin{\tB} \ g^{\mu \nu},  \\
      {\overline{\rho^-} \rho^- \HB} &: {i \frac{3}{2} \frac{\gB M_{\rho^-}}{\MZB}  \cos{\tB}},\\
      {\overline{\Psi^-} \Psi^- \HB} &: {i \frac{3}{2} \frac{\gB M_{\Psi^-}}{\MZB} \cos{\tB}}, \\
      {\overline{\Psi^-} \Psi^- h} &: {-i \frac{3}{2} \frac{\gB M_{\Psi^-}}{\MZB} \sin{\tB}}, \\
      \overline{\Psi^-} \Psi^- Z &: - i\frac{e \sin \theta_W}{\cos \theta_W} \gamma^\mu, \\
      \overline{\rho^-} \rho^- Z &:  i\frac{e \cos \theta_W}{\sin \theta_W} \gamma^\mu, \\
      \overline{\rho^-} \rho^0 W^- &: i g_2 \gamma^\mu, \\
      \overline{\Psi^-}e_i\phi_R &: \frac{i}{\sqrt{2}} \lambda_e^i P_R, \\
      \overline{\Psi^-}e_i\phi_I &: \frac{-i}{\sqrt{2}} \lambda_e^i P_R.
\end{align}
\newpage
\section{$h (p) \ZB (p_1) \gamma (p_2)$ Effective Interaction}
\label{hZBgamma}
The effective coupling between $h$, \ZB and the photon is generated at one-loop level, where inside the loop one has the top quark and computed using Package-X ~\cite{Patel:2015tea,Patel:2016fam}.
This coupling can be written as
\begin{equation}
\delta \Gamma^{\mu \nu}_{h \ZB \gamma}= \gB \frac{M_t}{v_0}\frac{e Q_t}{16 \pi^2} \left(A \ g^{\mu \nu} + B \ p_1^\mu p_2^\nu + C \ p_2^\mu p_1^\nu + D \ p_1^\mu p_1^\nu + E \ p_2^\mu p_2^\nu \right).
\end{equation}
Here $p_1^2=\MZB^2$ and $p_2^2=0$, being $p_1$ and $p_2$ the four momentum vectors associated to the \ZB and the photon, respectively. The coefficients in the above equations are given by
\begin{eqnarray}
    A &=& 4 M_t ( -2 + \frac{2 \MZB^2}{\MZB^2 -M_h^2} 
    ( \Lambda(M_h^2,M_t,M_t) - \Lambda (\MZB^2, M_t, M_t) ) \nonumber \\
    &+& (M_h^2 - 4 M_t^2 - \MZB^2) C_0 (0,M_h^2, \MZB^2, M_t, M_t, M_t)), 
    \nonumber \\
    B &=& - \frac{8 M_t}{(s - \MZB^2)^2} ( 2 \left(2 \MZB^2 + M_h^2 \right) \Lambda (\MZB^2, M_t, M_t) - 2 \left( \MZB^2 + 2 M_h^2 \right) \Lambda(M_h^2,M_t,M_t) \nonumber \\
    &+& (\MZB^2 - M_h^2) (6+ (4 M_t^2 
    + \MZB^2 + M_h^2) C_0 (0,\MZB^2, M_h^2, M_t, M_t, M_t) ), \nonumber
    \\
    C &=& \frac{2 A}{\MZB^2-M_h^2} , \nonumber \\
     D&=& 0, \nonumber \\
     E &=& - \frac{16 M_t \MZB^2}{(\MZB^2 - 
    M_h^2)^3}  (2 (2 \MZB^2 + M_h^2) ( \Lambda(\MZB^2, M_t, M_t) - 
      \Lambda (M_h^2, M_t, 
       M_t) ) \nonumber \\
       &+& ( \MZB^2 - 
        M_h^2) (6 + (4 M_t^2 + \MZB^2 + M_h^2) C_0(0, \MZB^2, M_h^2, M_t, M_t, 
          M_t))).
\end{eqnarray}
In the above equations the explicit form of the loop functions are:
\begin{eqnarray}
\Lambda (s,M_t,M_t)&=&\frac{\sqrt{s \left(s-4 M_t^2 \right)} \log \left(\frac{\sqrt{s \left(s-4 M_t^2\right)}+2 M_t^2-s}{2 M_t^2}\right)}{s}, \\
C_0 (0, M_h^2, \MZB^2, M_t,M_t,M_t) &=& \frac{\log ^2\left(\frac{- M_h^2+\sqrt{M_h^4-4 M_h^2 M_t^2}+2 M_t^2}{2 M_t^2}\right)-\log ^2\left(\frac{\sqrt{\MZB^4-4 M_t^2 \MZB^2}+2 M_t^2- \MZB^2}{2 M_t^2}\right)}{2 \left(M_h^2-\MZB^2\right)}. \nonumber \\
\end{eqnarray}

\bibliography{refs}

\end{document}